\documentclass[twocolumn,showpacs,preprintnumbers,amsmath,amssymb,floatfix]{revtex4}
\usepackage{graphicx}% Include figure files
\usepackage{dcolumn}% Align table columns on decimal point
\usepackage{bm}% bold math

\usepackage[usenames]{color}

\begin{document}

\preprint{ICCG14-0416}

\title{
Drift-Induced Step Instabilities Due to the Gap in the 
Diffusion Coefficient
}

\author{Masahide Sato}
	\email{sato@cs.s.kanazawa-u.ac.jp, 
Fax +81-76-234-6912}
\author{Makio Uwaha$^a$}
\author{Yukio Saito$^b$}
\affiliation{
	  Information Media Center of Kanazawa University,
		Kakuma-cho, Kanazawa 920-1192, Japan\\
	$^a$Department of Physics, Nagoya University,
		Furo-cho, Chikusa-ku, Nagoya 464-8602, Japan \\
	$^b$Department of Physics, Keio University,
		3-14-1 Hiyoshi, Kohoku-ku, Yokohama 223-8522, Japan
}
\date{\today}

%%%%%%%%%%%%%%%%%%%%%%%%%%%%%%%%%%%%%%%%%%%%%%%%%%%%%%%%%%%%%%%%%%%%%%%%%%

%------------------------------------------------------
\begin{abstract}
On  a Si(111) vicinal face near the structural transition temperature, 
the $1 \times 1$ structure and the $7 \times 7$ structure 
coexist in a terrace:
the $1 \times 1$ structure is in the lower side of the step edge
and the $7 \times 7$ structure in the upper side.
The diffusion coefficient of adatoms is different in the two structures.
Taking account of the gap in the diffusion
 coefficient at the step,
we study the possibility of step wandering induced by drift of adatoms.
A linear stability analysis
shows that 
the step wandering always occurs with step-down drift 
if the diffusion coefficient has a gap at the step.
Formation of straight grooves  by the step wandering
is expected from a nonlinear
analysis. 
The stability analysis also shows that 
step bunching occurs irrespective of the drift direction
if the diffusion in the lower side of the step is faster. 
The step bunching disturbs the formation of grooves.
If step-step repulsion is strong,
however, the step bunching is suppressed and the straight grooves appear.
Monte Carlo simulation confirms these predictions.
\end{abstract}

%-------------------------------------------------------------------------------

\maketitle

%-------------------------------------------------------------------------------
\section{Introduction}
On  vicinal faces of Si(111)~\cite{Degawa-ntmy99,Degawa-tmmyw01ss,Minoda02jpsj}
and  Si(001)~\cite{Nielsen-pp01ss}, step wandering occurs at high 
temperatures when a specimen is heated by direct electric current.
The current direction to cause the step wandering 
is step-down on the 
Si(111) vicinal face~\cite{Degawa-ntmy99,Degawa-tmmyw01ss,Minoda02jpsj}
and step-up  on the  Si(001) vicinal
 face~\cite{Nielsen-pp01ss}.

The cause of the step wandering is  drift of adatoms by the 
current~\cite{Stoyanov-90jjap,Metois-hp99ss,Ichikawa-d92apl,Degawa-mty00ssl}.  
The drift is in the same direction as 
the current~\cite{Metois-hp99ss,Ichikawa-d92apl}.
If the step is impermeable~\cite{Sato-us00prb,Sato-ush02prb}, 
the step wandering occurs with step-down drift,
as in the Si(111) vicinal 
face~\cite{Degawa-ntmy99,Degawa-tmmyw01ss,Minoda02jpsj}.
If there is alternation of the anisotropy in the diffusion coefficient
on consecutive  
terraces~\cite{Sato-ush03},
as  in the Si(001) vicinal face~\cite{Nielsen-pp01ss}
the step wandering occurs with step-up drift.

On Si(111) surfaces, 
the $1 \times 1$ structure is reconstructed and the $7 \times 7$ structure
appears at low temperatures ($\le 860^\circ$C).
In a vicinal face near the transition temperature,
the $7 \times 7$ structure spreads from  the upper side of  the steps, 
and the two  structures coexist in a terrace.
Recently, Hibino and co-workers~\cite{Hibino} observed step wandering
near $860^\circ$~C during growth. Due to the in-phase step wandering,
grooves perpendicular  to the steps appear on the vicinal face.
Kato and co-workers~\cite{Kato-USH-surf02} studied  the step wandering 
theoretically. Focusing on the difference in diffusion coefficient
of the two structures, they showed  that  the step wandering occurs
in growth if the diffusion coefficient on the $1 \times 1$ structure is
larger than that on  the $7 \times 7$ structure.

With the two phases coexisting,
the drift of adatoms
may also cause the step wandering instability on the Si(111)
vicinal face.
In this paper, we study the possibility of morphological instabilities
induced by the drift of adatoms
with the gap in the diffusion coefficient on the upper and the lower terraces.

%-------------------------------------------------------------------------------
\section{Model}~\label{sec:model}
We consider a vicinal face 
where steps are running parallel to the $x$-direction
bordering terraces of a width $l$ on average.
The $y$-direction is chosen  toward the step-down direction.
If impingement and evaporation of adatoms are neglected,
the adatom density $c(\mbox{\boldmath $r$},t)$
is determined  by
	\begin{equation}
	\frac{\partial c(\mbox{\boldmath $r$},t)}{\partial t}
	= \nabla 
	\cdot
	\mbox{\boldmath $j$}(\mbox{\boldmath $r$},t),
	\label{eq:diffusion-equation}
	\end{equation}
where $\mbox{\boldmath $j$}(\mbox{\boldmath $r$},t)$
 is the adatom current on the surface.
With step-down drift, the adatom current is given by
	\begin{equation}
	 \mbox{\boldmath $j$}(\mbox{\boldmath $r$},t)
	= -
	D_\mathrm{s}(\mbox{\boldmath $r$}) 
	\left(
	\mbox{\boldmath $\nabla c$}(\mbox{\boldmath $r$},t)
	- \frac{ F c(\mbox{\boldmath $r$},t)  }{k_\mathrm{B} T}
	\ \mbox{\boldmath $\hat{e}$}_y
	\right),
	\label{eq:adatomcurrent}
	\end{equation}
where $D_\mathrm{s}(\mbox{\boldmath $r$})$ is the local 
diffusion coefficient,
$F$ the force to cause the drift 
and is positive for the step-down drift,
and $\mbox{\boldmath{$\hat{e}_y$}}$
the unit vector toward the step-down direction. 
We assume that the diffusion coefficient $D_\mathrm{s}$ takes two values
in a terrace: 
$D_\mathrm{s} =D_1$ in  the lower side of a 
step edge,
$y_n< y<    y_n+l_1^{(n)} $, 
and $D_\mathrm{s} =D_2 $ in the upper side,
$y_{n-1} + l_1^{(n-1)} < y< y_{n}$,
where 
$y_n(x,t)$ is the position of the $n$th step
and 
$l_1^{(n)}(x,t)$
 is the terrace width of the lower side structure
($1 \times 1$ in Si(111)).

Solidification  and melting occur at step edges.
In local equilibrium at a step,
the adatom density is given by
	\begin{equation}
	\left. c\right|_{y_n }
	= 
	c_\mathrm{eq}^0
	\left( 
	1
        + \frac{\Omega \tilde{\beta}}{k_\mathrm{B}T }
	 \kappa
	+
	\frac{\Omega }{k_\mathrm{B}T}\frac{\partial U_n}{\partial y_n}
	\right),
	\label{eq:boundary-step}
	\end{equation}
where
$c_\mathrm{eq}^0$ is the equilibrium adatom density at the  isolated step,
$\Omega$ the atomic area, 
$\tilde{\beta}$ the step stiffness,
$\kappa$  the step curvature 
and $U_n$ the step-step interaction potential.

From the continuity of the adatom current  and the adatom density,
the boundary conditions at the phase boundary
 are  given by 
	\begin{eqnarray}
	\mbox{\boldmath $\hat{n}$}_\mathrm{b} \cdot
	\left. \mbox{ \boldmath $j$} \right
	|_{(y_n+ l_1^{(n)})+} 
	&=& 
		\mbox{\boldmath $\hat{n}$}_\mathrm{b} \cdot
	\left. \mbox{\boldmath $j$} \right
	|_{(y_n+l_1^{(n)})-},
	\label{eq:boundary-terrace1}
	\\
	\left. 
	c
	\right|_{(y_n+l_1^{(n)})+} 
	&=&
 	\left. 
  c
	\right|_{(y_n+l_1^{(n)})-},
	\label{eq:boundary-terrace2}
	\end{eqnarray}
where $\mbox{\boldmath $\hat{n}$}_\mathrm{b}$ is the normal vector 
of the boundary  and $+(-)$ indicates the lower (upper) side of the step.

By solving the diffusion equation (\ref{eq:diffusion-equation})
in a static approximation
with the boundary conditions,
Eqs.~(\ref{eq:boundary-step})-(\ref{eq:boundary-terrace2}),
the adatom density is determined.
The  normal step velocity $V_n$ is given by
	\begin{equation}
	V_n= \Omega \mbox{\boldmath $\hat{n}$}_\mathrm{s} \cdot
	(\left. \mbox{\boldmath $j$}  \right|_{y_n-} 
	-\left. \mbox{\boldmath $j$}  \right|_{y_n+} ),
	\end{equation}
where $\mbox{\boldmath $\hat{n}$}_\mathrm{s}$ is the normal
vector of the step.

\section{Stability analysis}\label{sec:stability-analysis}

When the steps and the boundaries of two phases are straight, 
the adatom density  in the quasi-static approximation
is given by
$	 c(y) = A_0+ B_0 e^{f (y-y_n)},$
where $f = F/k_\mathrm{B} T$.
In the lower side of a step,
% ($ < y< l_1 + y_n$), 
the coefficients $A_0$ and $B_0$ are given by
	\begin{eqnarray}
	 A_0 
	&=& 
	\frac{D_2 c_\mathrm{eq}^0(e^{fl^{(n)}}-1)
	}{D_2 e^{fl_2}(e^{fl_1^{(n)}} -1) + D_1(e^{fl_2^{(n)}} -1)},
	\label{eq:A01}
	\\
	 B_0 
	&=& 
	\frac{(D_1- D_2 )c_\mathrm{eq}^0(e^{fl_2^{(n)}}-1)
	}{D_2 e^{fl_2^{(n)}}(e^{fl_1^{(n)}} -1) + D_1(e^{fl_2^{(n)}} -1)},
	\label{eq:B01}
	\end{eqnarray}
where  $l^{(n)} = y_{n+1} -y_n$ and 
$l_2^{(n)} =  l^{(n)}- l_1^{(n)}$.
In the upper side of the step,
% ($l_1 < y< l$), 
the coefficients are given by the same form with 
the replacement $D_1 \leftrightarrow D_2$
and $ l_1^{(n)} \leftrightarrow  l_2^{(n)}$.
The adatom current $j_0$ is constant on the whole 
of the $n$-th terrace
and  is given by
	\begin{equation}
	j_0(l^{(n)})=
	\frac{D_1 D_2 c_\mathrm{eq}^0 f (e^{fl^{(n)}}-1)
	}{D_2 e^{fl_2^{(n)}}(e^{fl_1^{(n)}} -1) + D_1(e^{fl_2^{(n)}} -1)}.
	\label{eq:j0}
	\end{equation}
Since we have neglected the step repulsion, 
the current is a function of $l_1^{(n)}$ and $l_2^{(n)}$
 and does not depend on the neighboring terrace widths.

If the ratio of the width of the two structures 
$\gamma = l_2^{(n)}/l_1^{(n)}$
 is fixed, 
Eq.(\ref{eq:j0}) gives the current $j_0(l)$ 
as a function of the terrace width.
When the steps are equidistant,
the adatom current at the step positions
from the upper terrace equals
to that onto the lower terrace,
$\left. j_0 \right|_{y_n-} =\left. j_0 \right|_{y_n+}$.
The velocity of the steps vanishes and the steps do not move. 
If the step interaction is neglected, the stability 
of the equidistant steps
for step pairing 
is determined by  $j^\prime(l)$
since it controls the balance of the incoming and outgoing current 
with a pairing fluctuation of the terrace width
(the repulsive interaction tends to stabilize the system).
It is unstable for step pairing 
if $j_0^\prime(l)>0$ and stable otherwise. 
From Eq.~(\ref{eq:j0}),
$j_0^\prime(l)$ is given by
	\begin{eqnarray}
%	\begin{equation}
	j_0^\prime(l)
	&=&
	\frac{(D_1-D_2)D_1 D_2 c_\mathrm{eq}^0 f^2 e^{f l_2} 
	l_1 l
%	g(l)
	}{l[D_2 e^{fl_2}(e^{fl_1} -1) 
+ D_1(e^{fl_2} -1)]^2}
	\nonumber \\
	& &
	\times 
	\left(
	 \frac{e^{fl}-1}{l}
	 -
	 \frac{e^{fl_1}-1}{l_1}
	\right).
	\label{eq:jprime}
%	\end{equation}
	\end{eqnarray}
%where 
%$g(l) =  l_1 l[ (e^{f l}-1)/l -( e^{f l_1}-1)/l_1 ]$.
When the surface diffusion in the lower side of the step is faster
than that in the upper side ($D_1 >D_2$),
the vicinal face is unstable for the step pairing.
The stability is independent of the ratio $\gamma$ of the widths 
and the drift direction.

%%%%%%%%%%%%%%%%%%%%%%%%%%%%%%%%%%%%%%%%%%%%%%%%%%%%%%%%%%%%%%%%%
\begin{figure}[htp]
\centerline{
\includegraphics[width=0.6\linewidth]{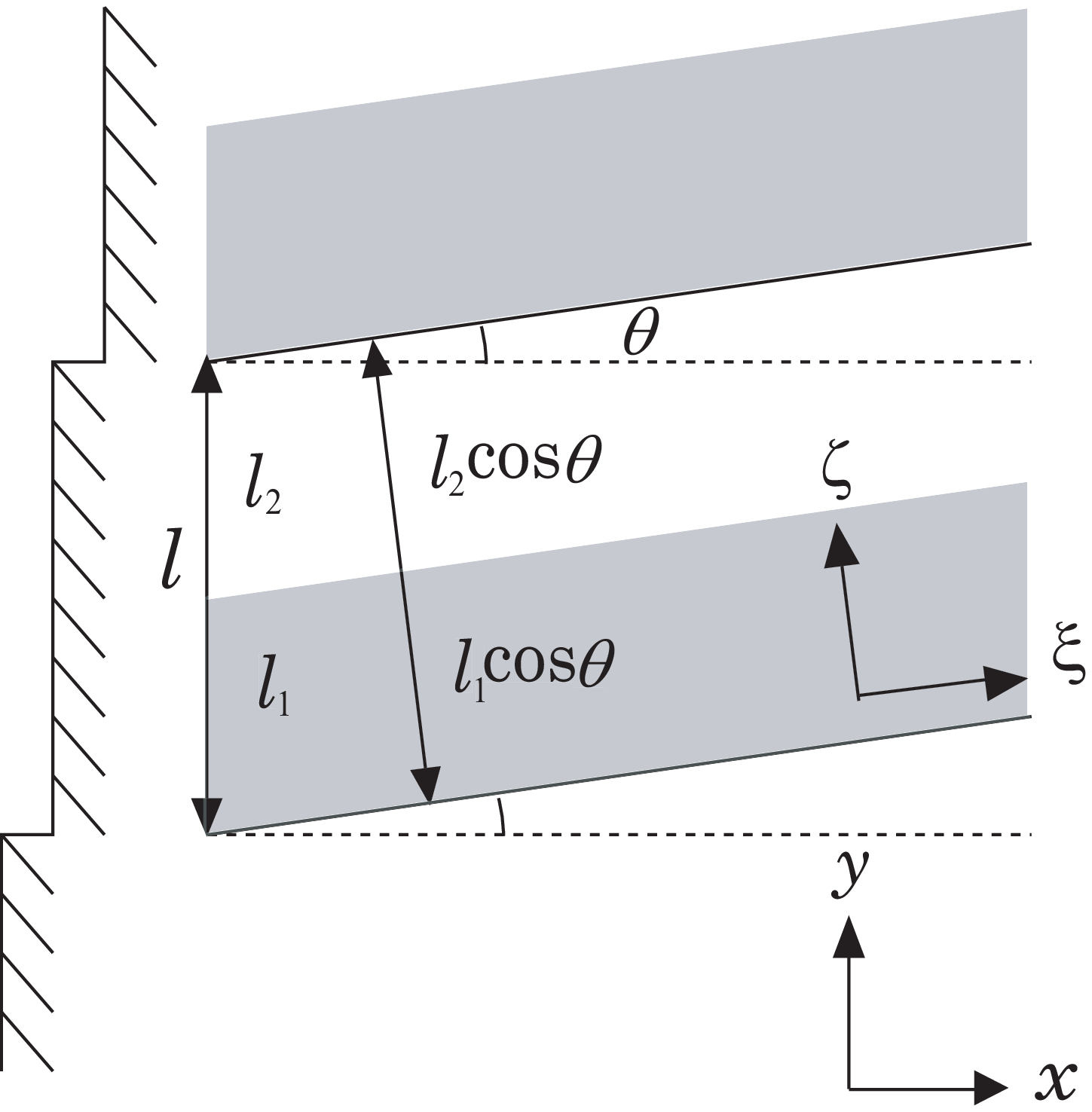}
}
\caption{A terrace bounded by tilted steps with an angle $\theta$.
}
\label{fig:tiltedsurface}
\end{figure}
%%%%%%%%%%%%%%%%%%%%%%%%%%%%%%%%%%%%%%%%%%%%%%%%%%%%%%%%%%%%%%%%%
To find the condition for the in-phase  step wandering,
we consider an equidistant train of straight steps tilted with 
an angle $\theta$ (Fig.~\ref{fig:tiltedsurface}).
The adatom current on the  terrace 
in the $x$-direction, $J_x^{(0)}$ is calculated as
	\begin{eqnarray}
	J_x^{(0)}
	&=& 
	\int_{y_n}^{y_{n+1}} dy   ( j_\parallel  \cos \theta
	-   j_\perp \sin \theta )
  \nonumber \\
	&=&
	\frac{
	(D_1-D_2 e^{f_\perp l_{1\perp}})
	(e^{f_\perp l_{1\perp} }-1)(e^{f_\perp l_{2\perp} }-1)
%(D_1-D_2) c_\mathrm{eq}^0 \tan \theta
	}{
	D_2 e^{fl_2}(e^{f_\perp l_{1\perp}} -1)
	 + D_1(e^{f_\perp l_{2\perp}} -1)
	} 
	\nonumber \\
	& &
	 \times 
	(D_1-D_2) c_\mathrm{eq}^0 \tan \theta,
	\end{eqnarray}
%%%%%%%%%%%%%%%%%%%%%%%%%%%%%%%%%%%%%%%%%%%%%%%%%%%%%%%%%%
%where $\parallel$ and $\perp$ indicate the $\xi $ and $\zeta$ directions.
where $j_\parallel$ and $j_\perp$ indicate the adatom current
in the $\xi $ and $\zeta$ directions,
$f_\perp = f \cos \theta$,
$l_{\perp} = l \cos \theta$,
$l_{1\perp} = l_1 \cos \theta$
and  $l_{2\perp} = l_2 \cos \theta$.
When the step distance is small enough, 
$f_\perp l_\perp \ll 1$,
it becomes
	\begin{eqnarray}
	J_x^{(0)}
	&=&
	\frac{(D_1 -D_2)^2 f l_1 l_2 c_\mathrm{eq}^0}{(D_1 l_1 + D_2 l_2)}
	\frac{\eta_x}{1+ (\eta_x)^2}.
	\label{eq:adatom-current1}
	\end{eqnarray}
where we assume  $y_n(x,t) = n l + \eta(x,t) $
and $\eta_x = \partial \eta/\partial x$.
%%%%%%%%%%%%%%%%%%%%%%%%%%%%%%%%%%%%%%%%%
In addition to Eq.~(\ref{eq:adatom-current1}),
there is current $J_x^{(1)}$ caused by the change of the chemical 
potential along the step:
	\begin{equation}
	 J_x^{(1)} 
	= - \cos^2 \theta (D_1 l_{1\perp} + D_2 l_{2\perp})c_\mathrm{eq}^0
	\frac{\partial }{\partial x}\left( \frac{\mu}{k_\mathrm{B}T} \right),
	\label{eq:adatom-current2}
	\end{equation}
where $\mu = \Omega \tilde{\beta} \kappa$. 

The evolution of the step position is determined by
the adatom current in the $x$-direction as
	\begin{eqnarray}
	\frac{\partial \eta}{\partial t}
	 &=&
	-\Omega \frac{\partial (J_x^{(0)} + J_x^{(1)})}{\partial x}
	\nonumber \\
	&=&
	-
	\frac{\partial }{\partial x}
	 \left[ 
	 \frac{ \alpha_2 \eta_x}{1+ \eta_x^2}
	 +
	 \frac{ \alpha_4 }{1+ \eta_x^2}
	 \frac{\partial }{\partial x}
	 \left( 
	 \frac{\eta_{xx}}{(1+ \eta_x^2)^{3/2}}
	 \right)
	\right], \ \ \ \
	\label{eq:nonlinear-equation}
	\end{eqnarray}
where the coefficients $\alpha_2$ and $\alpha_4$ are
	\begin{eqnarray}
	\alpha_2
	&=& 
	\Omega \frac{(D_1-D_2)^2 fl_1 l_2 c_\mathrm{eq}^0
	}{(D_1 l_2 + D_2 l_1)	},
	\label{eq:alpha2}
	\\
	\alpha_4
	&=&
	\Omega (D_1 l_1 + D_2 l_2) c_\mathrm{eq}^0
	\frac{\Omega \tilde{\beta}}{k_\mathrm{B}T}.
	\label{eq:evolution-equation}
	\end{eqnarray}

If the step position is  of the form $\eta (x) = \eta _0 e^{iqx + \omega_q t}$,
the linear amplification rate is
	\begin{equation}
	\omega_q = \alpha_2 q^2 - \alpha_4 q^4.
	\label{eq:amplification-rate}
	\end{equation}
The coefficient $\alpha_4 $ is always positive
and suppresses the step fluctuation.
With step-up drift ($ f <0$) the 
coefficient $\alpha_2$ is negative and suppresses the step fluctuation,
while  $\alpha_2$ becomes positive and 
the step wandering occurs
with step-down drift ($ f >0$).

Equation~(\ref{eq:nonlinear-equation})
is the same type of equation describing 
the step wandering in other conserved 
systems~\cite{Kato-USH-surf02,Pierre-Louis-mskp98prl,Gillet-pm00epjb,%
Politi-m04prl}.
The solution of the equation shows a regular 
periodic pattern whose amplitude 
increases in a power law of time as $t^{1/2}$~\cite{Pierre-Louis-mskp98prl}.
As a result periodic grooves will be formed.

\section{Monte Carlo simulation}\label{sec:mc}
We perform Monte Carlo simulation for solid-on-solid steps
of a square lattice model.
The boundary condition is helical in the $y$-direction
and periodic in the $x$-direction. 
We assume that $\gamma$ is fixed to 1 so that the phase boundary is
at $(y_n+y_{n+1})/2$ when the steps move.
In the lower side of a step (supposedly the $1\times 1$ region),
an adatom on the site $(i,j)$ moves to $(i\pm 1,j)$ with the probability
$p_\mathrm{d} = D_1/4$ and to 
$(i,j\pm 1)$ with the probability $p_\mathrm{d} =D_1(1 \pm fa/2)/4$.
In the upper side of a step (the $7\times 7$ region), 
the parameter $D_1$  is replaced by $D_2$.
The diffusion across the boundary of the two  regions takes  $D_2$.
In our simulations, 
we assume that  
$(D_1, D_2) =(1, \alpha)$ or
$(\alpha,1)$ with  $\alpha <1$.

Solidification and melting occur 
at the lower edge of the step positions~\cite{Saito-u94prb}.
The  probabilities  for solidification $p_+$ and melting $p_-$
are given by
	\begin{equation}
	\label{eq:solidification}
	p_{\pm}
	=
	\left [
	1 
	+ 
	{\rm exp} 
	\left ( \frac{\Delta E_\mathrm{s}  + \Delta U \mp \phi}{k_{\rm B} T} \right)
	\right ]^{-1} ,
	\end{equation}
where $\Delta E_\mathrm{s}$ is the increment of the step energy,
$\phi$ the potential gain by solidification
and 
$\Delta U$ is the change of the step-step interaction potential.
We assume the repulsive interaction potential $U_n$  of the $n$th step
takes the form 
\begin{equation} 
U_n = \sum_{m=n\pm 1 }
	 \frac{A}{[y_n(x_i) -y_{m}(x_i)]^{2}},
\end{equation}
where $y_n(x_i)$ is the position of the $n$th step at $x=x_i$.

We first carry out the simulation with the diffusion coefficients
$(D_1, D_2) = (1, 0.1)$,
the stiffness $\tilde{\beta}/k_\mathrm{B}T=1.64$
and the equilibrium density $c_\mathrm{eq}^0  = 0.18$.
The system size is $ 512 \times 512$ and the initial step distance is $l=16$
(the step number is $N=32$).
Initially,
the steps  are  straight and equidistant, and 
there are a few adatoms on the terraces.

%%%%%%%%%%%%%%%%%%%%%%%%%%%%%%%%%%%%%%%%%%%%%%%%%%%%%%%%%%%%%%%%%
\begin{figure}[htp]
\centerline{
\includegraphics[width=0.6\linewidth]{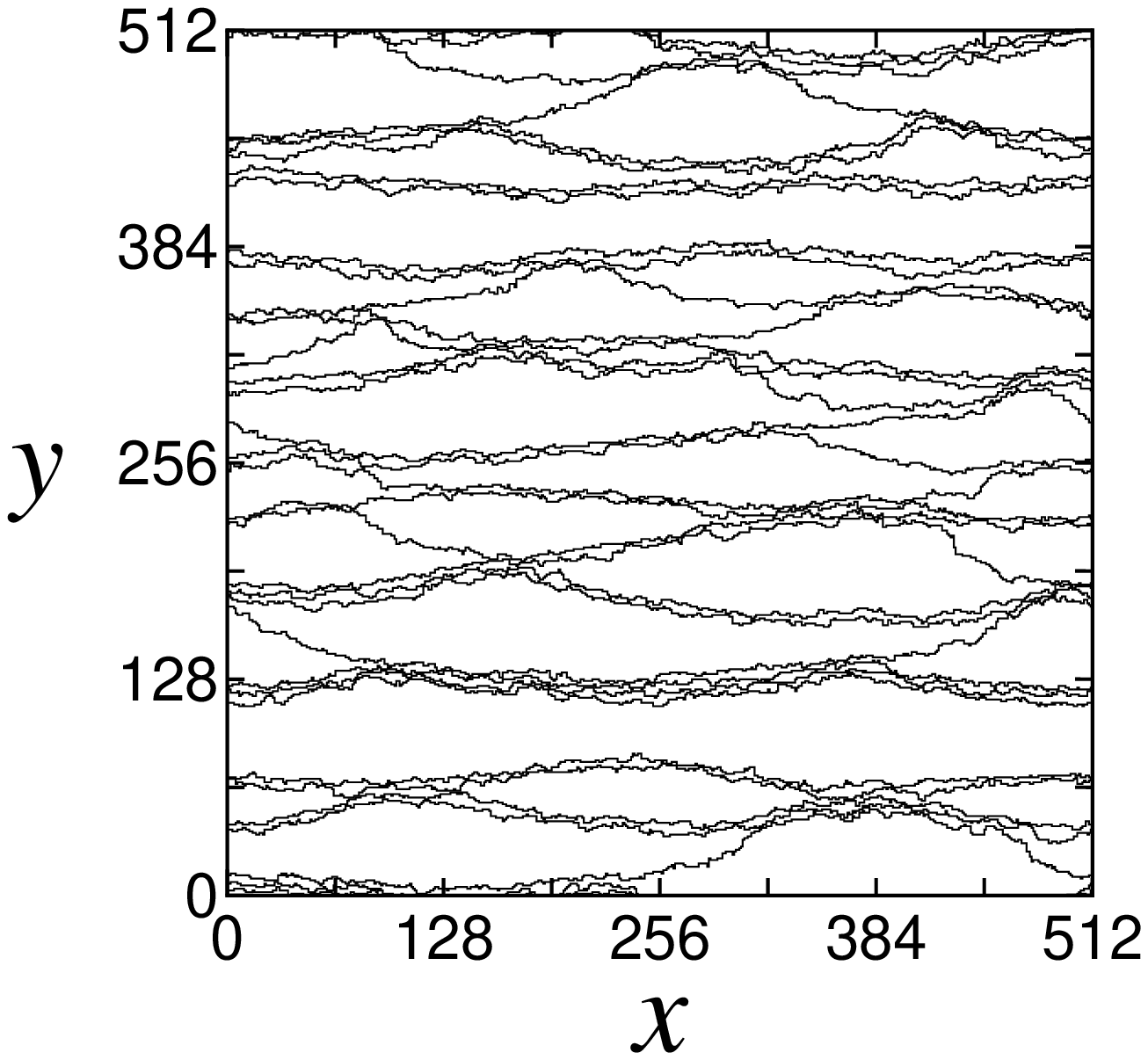}
}
\centerline{\sf \Large \hspace*{0.3cm} (a) }

\centerline{
\includegraphics[width=0.6\linewidth]{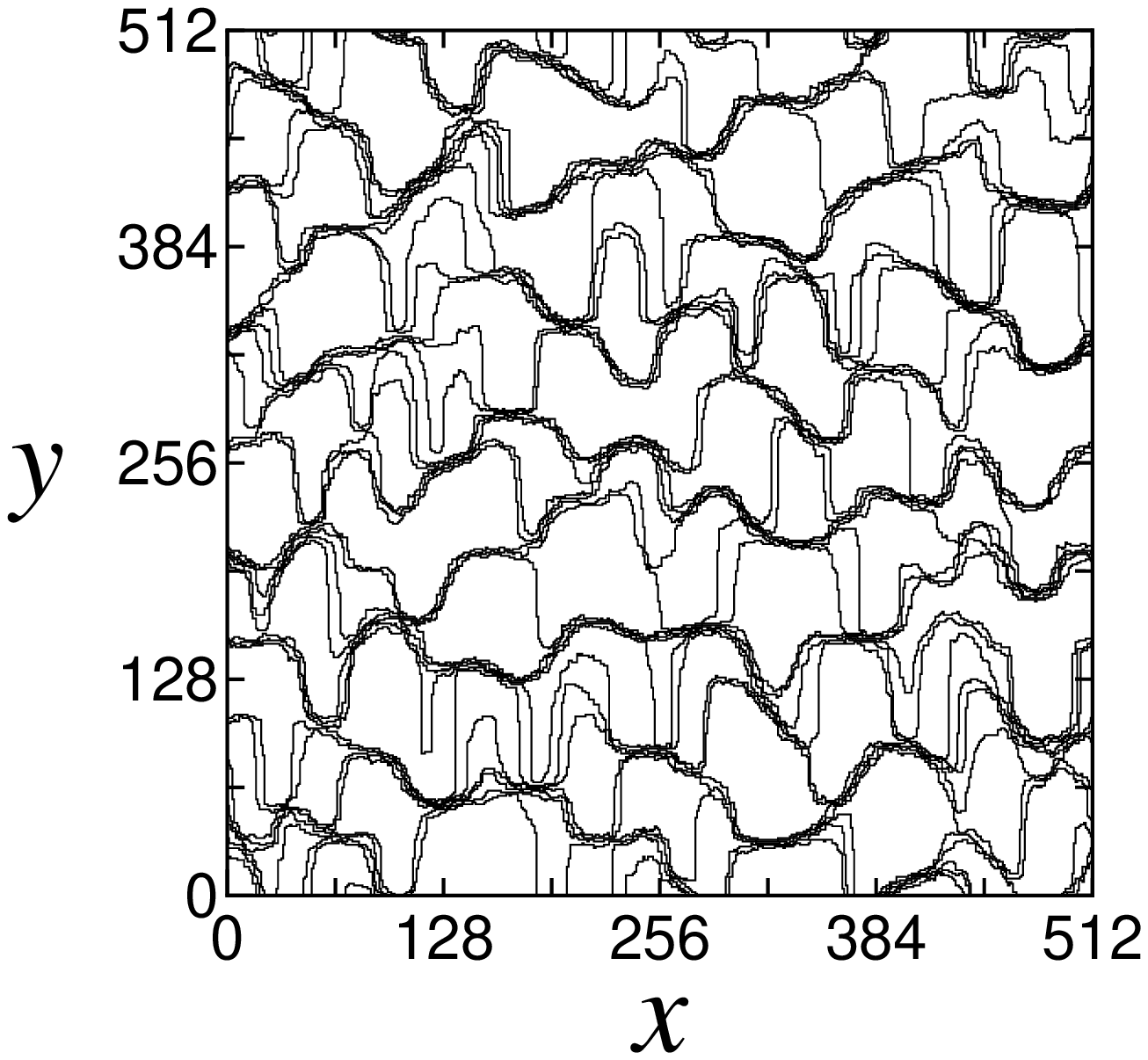}
}
\centerline{\sf \Large  \hspace*{0.3cm} (b)}
\caption{Snapshots of a destabilized vicinal face for $D_1 >D_2$:
with (a) step-up drift
and (b) step-down drift
.}
\label{fig:snapshot}
\end{figure}
%%%%%%%%%%%%%%%%%%%%%%%%%%%%%%%%%%%%%%%%%%%%%%%%%%%%%%%%%%%%%%%%%

%%%%%%%%%%%%%%%%%%%%%%%%%%%%%%%%%%%%%%%%%%%%%%%%%%%%%%%%%%%%%%
\begin{figure}[htp]
\centerline{
\includegraphics[width=0.6\linewidth]{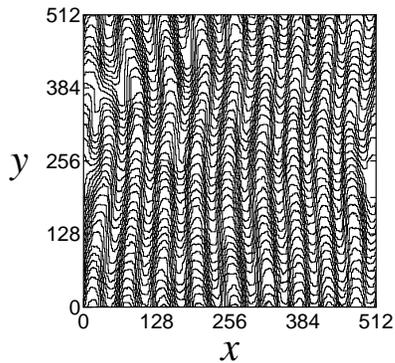}
}
\caption{
Snapshot of step wandering induced by step-down drift
with a strong step-step repulsion for $D_1 >D_2$.
%The step-step repulsive potential is $U=A/l^2$ with $A =360$.
Except for the repulsive potential,
the  parameters are the same as in Fig.~\ref{fig:snapshot}(b).
}
\label{fig:snapshot1}
\end{figure}
%%%%%%%%%%%%%%%%%%%%%%%%%%%%%%%%%%%%%%%%%%%%%%%%%%%%%%%%%%%%%%

When we neglect the step-step repulsive interaction,
the vicinal face is unstable for the step bunching (Fig.~\ref{fig:snapshot}(a)),which agrees with the analysis in Sec.~\ref{sec:stability-analysis}.
The step bunching occurs irrespective of the drift direction,
but the form of the bunches changes with the drift direction.
The bunches are straight with step-up drift (Fig.~\ref{fig:snapshot}(a))
and wander with step-down drift (Fig.~\ref{fig:snapshot}(b)).
With a  strong repulsive interaction, $A = 300$,
the step bunching is suppressed
and the in-phase step wandering occurs.
%%%%%%%%%%%%%%%%%%%%%%%%%%%%%%%%%%%%%%%%
%Grooves parallel to the drift are produced  (Fig.~\ref{fig:snapshot1}),
%which  agrees with the 
%solution~\cite{Gillet-pm00epjb,Pierre-Louis-mskp98prl} 
%of Eq.~(\ref{eq:nonlinear-equation}).
We have not studied the growth laws of
the step width and 
the period of grooves, but the form of the grooves
(Fig.~\ref{fig:snapshot1})
qualitatively agrees with the 
solution~\cite{Gillet-pm00epjb,Pierre-Louis-mskp98prl} 
of Eq.~(\ref{eq:nonlinear-equation}).

%%%%%%%%%%%%%%%%%%%%%%%%%%%%%%%%%%%%%%%%%%%%%%%%%%%%%%%%%%%%%%%%%
\begin{figure}[htp]
\vspace*{0.2cm}
\centerline{
\includegraphics[width=0.6\linewidth]{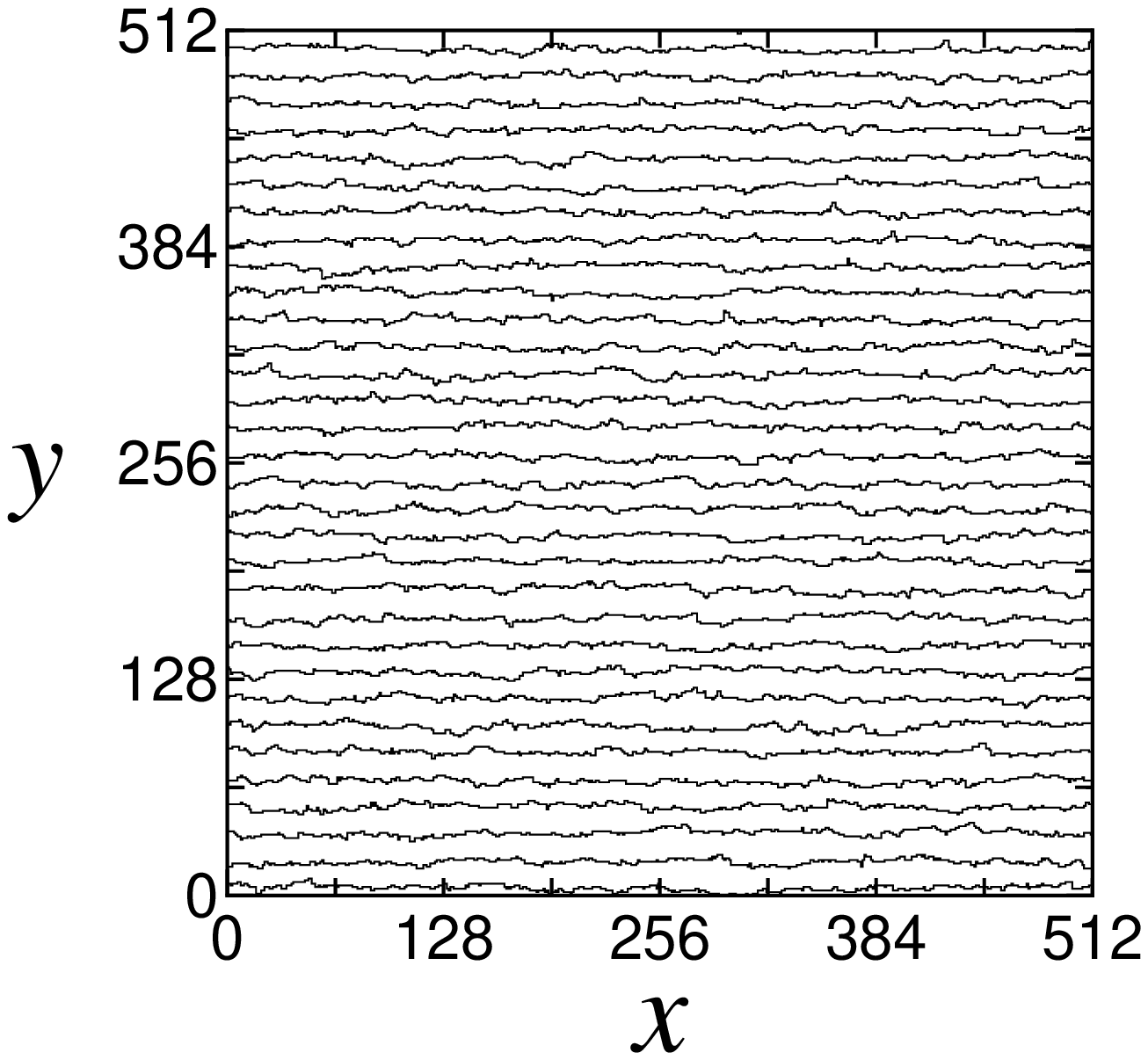}
}
\centerline{\sf \Large \hspace*{0.3cm} (a) }
\centerline{
\includegraphics[width=0.6\linewidth]{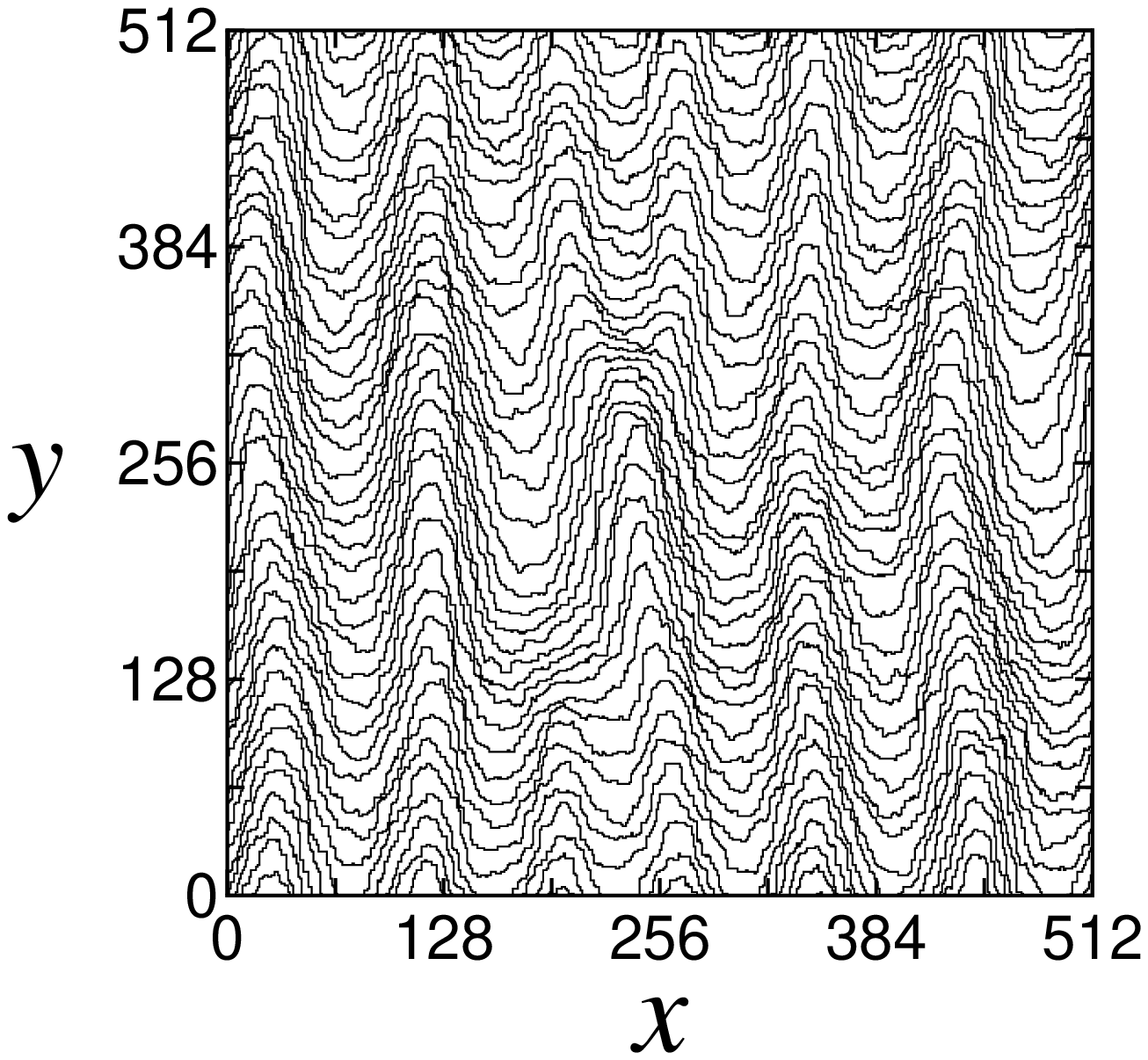}
}
\centerline{\sf \Large \hspace*{0.3cm}(b)}
\caption{Snapshots of a  vicinal face for $D_1 <D_2$:
with (a) step-up drift
and 
(b) step-down drift.
The repulsive interaction is absent.
}
\label{fig:snapshot2}
\end{figure}
%%%%%%%%%%%%%%%%%%%%%%%%%%%%%%%%%%%%%%%%%%%%%%%%%%%%%%%%%%%%%%%%%

We also carry out the simulation with $D_1 < D_2$
 (Fig.~\ref{fig:snapshot2}).
The diffusion coefficients are 
$(D_1, D_2) = (0.1,1 )$
and the step-step repulsion is neglected ($A=0$).
Other parameters 
are the same as  Fig.~\ref{fig:snapshot}.
With step-up drift (Fig.~\ref{fig:snapshot2}(a)), 
neither step wandering nor step bunching occurs.
With step-down drift,
the step wandering occurs,
but no indication of bunching is seen (Fig.~\ref{fig:snapshot2}(b)).
As time passes,
%the  disturbed wandering pattern in the initial stage
%is spontaneously eliminated, 
periodicity selection proceeds,
resulting in straight grooves 
as those of Fig.~\ref{fig:snapshot1}.

\section{Summary and discussion}\label{sec:summary}
In this paper, we studied the drift-induced morphological instabilities
on a vicinal face with two phases.
As is seen from Eq.~(\ref{eq:alpha2}), 
the step wandering occurs with step-down drift unless $D_1=D_2$. 
If $D_1=D_2$, our model reduces to a model of a simple vicinal face. 
We have already found~\cite{Sato-ush02prb} that, with step-down drift, 
wandering instability occurs in such a simple vicinal face 
if the kinetic coefficient of the step is finite. 
Although in the present paper we have assumed an infinite kinetic 
coefficient (local equilibrium at the steps), 
it must be finite in reality. 
Therefore we may say wandering instability is 
expected irrespective of the diffusion ratio. 
In all cases
straight grooves parallel to the drift
is produced due to the in-phase step wandering in 
Monte Carlo simulation if step bunching is suppressed.

In the Si(111) vicinal face near the transition temperature,
the  diffusion coefficient in the lower side of a step 
is  larger than that in the upper side~\cite{Hibino-HuOT}.
With the assumption that the ratio of the widths of the 
two structures is constant,
the step bunching occurs irrespective of the current direction 
and the step wandering occurs with step-down current.
In a vicinal face of large inclination, however,
%However, if the inclination of the vicinal face is large,
the step bunching is suppressed due to 
the strong step-step repulsion,
and only grooves induced by the step wandering may be observed
with step-down current.

In our model,
we assumed that the boundary of the two structures moves 
in concert with the 
steps and 
the ratio of the widths of two structures is constant.
The drift direction to cause the instabilities does not
depend on the ratio of the widths.
In reality the ratio of the widths
changes 
with temperature~\cite{Yamaguchi-y93ss} and the motion of the boundary
 does not automatically follow the steps.
To study the morphological development of this system in detail,
we need to extend our model to include the freedom of the motion of the
 boundary~\cite{kato-us04ss}.

\acknowledgements
This work was supported by Grant-in-Aid for Scientific
Research from Japan Society  for the Promotion of Science.
M. U. and Y. S. benefited from the inter-university cooperative
research program of the Institute for Materials Research, Tohoku University.

%%%%%%%%%%%%%%%%%%%%%%%%%%%%%%%%%%%%%%%%%%%%%%%%%%%%%%%%%%%%%

\end{document}